\documentclass[12pt]{article}
\usepackage[english]{babel}
\usepackage{amsmath,amsthm,amssymb,amsfonts,url}
\usepackage{color,graphicx}

\newtheorem{definition}{Definition}

\newtheorem{example}{Example}

\allowdisplaybreaks
\begin{document}

\title{Approximate Q--conditional symmetries\\
of partial differential equations}

\author{M.~Gorgone and F.~Oliveri\\
\ \\
{\footnotesize Department of Mathematical and Computer Sciences,}\\
{\footnotesize Physical Sciences and Earth Sciences, University of Messina}\\
{\footnotesize Viale F. Stagno d'Alcontres 31, 98166 Messina, Italy}\\
{\footnotesize mgorgone@unime.it; foliveri@unime.it}
}

\date{Published in \textit{Electronic Journal of Differential Equations} \textbf{25}, 133--147 (2018).}
\maketitle


\begin{abstract}
Following a recently introduced approach to approximate Lie symmetries of differential equations which is 
consistent with the principles of perturbative analysis of differential equations containing small terms, we
analyze the case of approximate $Q$--conditional symmetries. An application of the method to a hyperbolic variant of a 
reaction--diffusion--convection equation is presented.
\end{abstract}


\maketitle
\bigskip
\centerline{\emph{Dedicated to the memory of Anna Aloe}}

\section{Introduction}
Lie continuous groups of transformations, originally introduced in the nineteenth century by Sophus Lie \cite{Lie1,Lie2}, provide a unified and algorithmic approach to the investigation of ordinary and partial differential equations.
Lie continuous transformations, completely characterized by their infinitesimal generators, establish a diffeomorphism on the space of independent and dependent variables, taking solutions of the equations into other solutions (see the monographies \cite{Baumann,BlumanAnco,BlumanCheviakovAnco,Bordag,Ibragimov,Ibragimov:CRC,Meleshko2005,Olver1,Olver2,Ovsiannikov}, or the review paper \cite{Oliveri:Symmetry}). Any transformation of the independent and dependent variables in turn induces a transformation of the derivatives. As Lie himself showed,  the problem of finding the Lie group of point transformations leaving a differential equation invariant consists in solving a \emph{linear} system of determining equations for the components of its infinitesimal generators. Lie's theory allows for  the development of systematic procedures leading to the integration by quadrature (or at least to lowering the order) of ordinary differential equations, to the determination of invariant solutions of initial and boundary value problems, to the derivation of conserved quantities, or to the construction of relations between different differential equations that turn out to be equivalent \cite{BlumanAnco,BlumanCheviakovAnco,DonatoOliveri1994,DonatoOliveri1995,DonatoOliveri1996,GorgoneOliveri1,GorgoneOliveri2,Oliveri:Symmetry,Oliveri:quasilinear}.

Along the years, many extensions and generalizations of Lie's approach have been introduced, and a wide set 
of applied problems has been deeply analyzed and solved. Here, we limit ourselves to consider conditional symmetries whose origin can be thought of as originated in 1969 with the paper by Bluman and Cole 
\cite{Bluman-Cole-1969} on the \emph{nonclassical method}, where the continuous symmetries mapping the manifold characterized by the linear heat equation and the invariant surface condition (together with its differential consequences) into itself has been considered. Later on, 
Fuschich \cite{Fuschich,Fuschich-Tsyfra} introduced the general concept of conditional invariance. The basic idea of Q--conditional symmetries consists in replacing the conditions for the invariance of the given system of differential equations by the weak conditions for the invariance of the combined system made by the original equations, the invariant surface conditions and their differential consequences. As a result, fewer determining equations, which as an additional complication turn out to be nonlinear, arise, and in principle a larger set of symmetries can be recovered. Even if the nonlinear determining equations rarely can be solved in their generality, it is possible to find some particular solutions allowing us to find 
nontrivial solutions of the differential equations we are considering. The available literature on this subject is
quite large  (see, for instance, \cite{Arrigo,Clarkson-Kruskal,Clarkson-Mansfield,Levi-Winternitz,Nucci1993,NucciClarkson1992,Olver-Rosenau-1,Olver-Rosenau-2,Saccomandi});  many recent applications of Q--conditional symmetries to reaction--diffusion equations can be found in \cite{Cherniha-JMAA2007,Cherniha-JPA2010,ChernihaDavydovych-MCM2011,ChernihaDavydovych-CNSNS2012,ChernihaPliukhin-JPA2008}. 

In concrete problems, differential equations may contain terms of different orders of magnitude; the occurrence of small terms has the effect of destroying many useful symmetries,  and this restricts the set of invariant solutions that can be found.
One has such a limitation also when one looks for conditional symmetries.  To overcome this inconvenient, some \emph{approximate symmetry theories} have been proposed in order to deal with differential equations involving small terms, and the notion of  \emph{approximate invariance} has been introduced. The first approach was introduced by Baikov, Gazizov and Ibragimov \cite{BGI-1989} (see also \cite{IbragimovKovalev}), 
who in 1988 proposed to expand in a perturbation series the Lie generator in order to have
an approximate generator; this approach has been applied to many physical situations 
\cite{BaikovKordyukova2003,DolapciPakdemirli2004,GazizovIbragimovLukashchuk2010,GazizovIbragimov2014,Gan_Qu_ND2010,%
IbragimovUnalJogreus2004,Kara_ND2008,Kovalev_ND2000,Pakdemirli2004,Wiltshire1996,Wiltshire2006}. 
Though the theory is quite elegant, nevertheless, the expanded generator is not consistent with the principles of perturbation analysis 
\cite{Nayfeh} since the dependent variables are not expanded. This implies that in several examples the approximate invariant solutions that are found with this method are not the most general ones. A different approach has been proposed in 1989 by Fushchich and Shtelen \cite{FS-1989}: the dependent variables are expanded in a  series as done in usual perturbation analysis,
terms are then separated at each order of approximation, and a system of equations to be solved in a hierarchy is obtained.  This resulting system is assumed to be coupled, and the approximate symmetries of the original equation are defined as the \emph{exact symmetries} of the equations obtained from perturbations.
This approach, having an obvious simple and coherent basis, requires a lot of algebra; applications of this method to various equations can be found, for instance, in the papers 
\cite{Diatta,Euler1,Euler2,Euler3,Wiltshire2006}. In 2004, Pakdemirli \emph{et al.} \cite{Pakdemirli2004} compared the two different approaches, and proposed a third method as a variation of Fushchich--Shtelen one by removing 
the assumption of a fully coupled system. Another variant of Fushchich--Shtelen method has been proposed in \cite{Valenti}.

In a recent paper \cite{DSGO-lieapprox}, an approximate symmetry theory which is consistent with perturbation analysis, and such that the relevant properties of exact Lie symmetries of differential equations are inherited, has been proposed. More precisely, the dependent variables are expanded in power series of the small parameter as done in classical perturbation analysis; then, instead of considering the approximate symmetries as the exact symmetries of the approximate system (as done in Fushchich--Shtelen method), the consequent expansion of the Lie generator is constructed, and the approximate invariance  with respect to an approximate Lie generator is introduced, as in Baikov--Gazizov--Ibragimov method. An application of the method to the equations of a creeping flow of a second grade fluid can be found in 
\cite{Gorgone-IJNLM2018}.

For differential equations containing small terms, approximate Q--conditional symmetries can be considered as well. Within the theoretical framework proposed by Baikov, Gazizov and Ibragimov for approximate symmetries, approximate conditional symmetries \cite{Mahomed2000,Shih2005} have been considered.

In this paper, we aim to define approximate Q--conditional symmetries along the consistent approach developed in \cite{DSGO-lieapprox}. The plan of the paper is as follows. In Section~\ref{sec2}, after fixing the notation, we review the approach of approximate symmetries introduced in \cite{DSGO-lieapprox}. Then, in Section \ref{sec3}, we define approximate Q--conditional symmetries of differential equations. In Section~\ref{sec4}, an application of the method is given by considering a hyperbolic variant of a reaction--diffusion--convection equation. Finally, Section~\ref{sec5} contains  our conclusions.

\section{Lie symmetries and approximate Lie symmetries}
\label{sec2}
Let us consider an $r$th order differential equation
\begin{equation}
\label{de}
 \Delta\left(\mathbf{x},\mathbf{u},\mathbf{u}^{(r)}\right)=0,
\end{equation}
where $\mathbf{x}\equiv(x_1,\ldots,x_n)\in \mathcal{X}\subseteq \mathbb{R}^n$ are the independent variables, $\mathbf{u}\equiv(u_1,\ldots,u_m)\in \mathcal{U}\subseteq \mathbb{R}^m$ the dependent variables, and $\mathbf{u}^{(r)}$ denotes the derivatives up to the order $r$ of the $\mathbf{u}$'s with respect to the $\mathbf{x}$'s.

A Lie point symmetry of \eqref{de} is characterized by the infinitesimal operator
\begin{equation}
\label{op}
\Xi=\sum_{i=1}^n\xi_i(\mathbf{x},\mathbf{u})\frac{\partial}{\partial x_i}
+\sum_{\alpha=1}^m\eta_\alpha(\mathbf{x},\mathbf{u})\frac{\partial}{\partial u_\alpha}
\end{equation}
such that
\begin{equation}
\label{invcond}
\left.\Xi^{(r)}\left(\Delta\left(\mathbf{x},\mathbf{u},\mathbf{u}^{(r)}\right)\right)\right|_{\Delta=0}=0,
\end{equation}
where $\Xi^{(r)}$ is the $rth$--prolongation of \eqref{op} 
\cite{Baumann,BlumanAnco,BlumanCheviakovAnco,Bordag,Ibragimov,Ibragimov:CRC,Meleshko2005,Oliveri:Symmetry,Olver1,Olver2,Ovsiannikov}. Condition \eqref{invcond} leads to 
a system of linear partial differential equations  (determining equations) whose integration provides the \emph{infinitesimals} $\xi_i$ and $\eta_\alpha$. Invariant solutions corresponding to a given Lie symmetry are found by solving 
the invariant surface conditions
\begin{equation}
\label{Qconstraint}
\mathbf{Q}\equiv \sum_{i=1}^n \xi_i(\mathbf{x},\mathbf{u}) \frac{\partial\mathbf{u}}{\partial x_i}-
{\boldsymbol \eta}(\mathbf{x},\mathbf{u})=\mathbf{0},
\end{equation}
and inserting their solutions in \eqref{de}.

Now, let us consider differential equations containing small terms, say
\begin{equation}
\Delta\left(\mathbf{x},\mathbf{u},\mathbf{u}^{(r)};\varepsilon\right)=0,
\end{equation}
where $\varepsilon$ is a \emph{small constant parameter}, and review approximate symmetries following the approach described in \cite{DSGO-lieapprox}.

In perturbation theory \cite{Nayfeh}, a differential equation involving small terms is often studied by looking for solutions in the form
\begin{equation}
\label{expansion_u}
\mathbf{u}(\mathbf{x},\varepsilon)=\sum_{k=0}^p\varepsilon^k \mathbf{u}_{(k)}(\mathbf{x})+O(\varepsilon^{p+1}),
\end{equation}
whereupon the differential equation writes as
\begin{equation}
\Delta\approx \sum_{k=0}^p\varepsilon^k\widetilde{\Delta}_{(k)}\left(\mathbf{x},\mathbf{u}_{(0)},\mathbf{u}^{(r)}_{(0)},
\ldots,\mathbf{u}_{(k)},\mathbf{u}^{(r)}_{(k)}\right)=0.
\end{equation}
Now, let us consider a Lie generator
\begin{equation}
\Xi=\sum_{i=1}^n\xi_i(\mathbf{x},\mathbf{u};\varepsilon)\frac{\partial}{\partial x_i}
+\sum_{\alpha=1}^m\eta_\alpha(\mathbf{x},\mathbf{u};\varepsilon)\frac{\partial}{\partial u_\alpha},
\end{equation}
where we assume that the infinitesimals  depend on the small parameter $\varepsilon$.

By using the expansion~\eqref{expansion_u} of the dependent variables only, we have for the infinitesimals
\begin{equation}
\xi_i\approx\sum_{k=0}^p\varepsilon^k \widetilde{\xi}_{(k)i}, \qquad \eta_\alpha\approx\sum_{k=0}^p\varepsilon^k\widetilde{\eta}_{(k)\alpha},
\end{equation}
where
\begin{equation}
\begin{aligned}
&\widetilde{\xi}_{(0)i}=\xi_{(0)i}=\left.\xi_i(\mathbf{x},\mathbf{u},\varepsilon)\right|_{\varepsilon=0},\qquad
&&\widetilde{\eta}_{(0)\alpha}=\eta_{(0)\alpha}=\left.\eta_\alpha(\mathbf{x},\mathbf{u},\varepsilon)\right|_{\varepsilon=0,}\\
&\widetilde{\xi}_{(k+1)i}=\frac{1}{k+1}\mathcal{R}[\widetilde{\xi}_{(k)i}],\qquad &&\widetilde{\eta}_{(k+1)\alpha}=\frac{1}{k+1}\mathcal{R}[\widetilde{\eta}_{(k)\alpha}],
\end{aligned}
\end{equation}
$\mathcal{R}$ being a \emph{linear} recursion operator satisfying \emph{product rule} of derivatives and such that
\begin{equation}
\label{R_operator}
\begin{aligned}
&\mathcal{R}\left[\frac{\partial^{|\tau|}{f}_{(k)}(\mathbf{x},\mathbf{u}_{(0)})}{\partial u_{(0)1}^{\tau_1}\dots\partial u_{(0)m}^{\tau_m}}\right]=\frac{\partial^{|\tau|}{f}_{(k+1)}(\mathbf{x},\mathbf{u}_{(0)})}{\partial u_{(0)1}^{\tau_1}\dots\partial u_{(0)m}^{\tau_m}}\\
&\phantom{\mathcal{R}\left[\frac{\partial^{|\tau|}{f}_{(k)}(\mathbf{x},\mathbf{u}_{(0)})}{\partial u_{(0)1}^{\tau_1}\dots\partial u_{(0)m}^{\tau_m}}\right]}
+\sum_{i=1}^m\frac{\partial}{\partial u_{(0)i}}\left(\frac{\partial^{|\tau|} {f}_{(k)}(\mathbf{x},\mathbf{u}_{(0)})}{\partial u_{(0)1}^{\tau_1}\dots\partial u_{(0)m}^{\tau_m}}\right)u_{(1)i},\\
&\mathcal{R}[u_{(k)j}]=(k+1)u_{(k+1)j},
\end{aligned}
\end{equation}
where $k\ge 0$,  $j=1,\ldots,m$, $|\tau|=\tau_1+\cdots+\tau_m$.

Thence, we have an approximate Lie generator
\begin{equation}
\Xi\approx \sum_{k=0}^p\varepsilon^k\widetilde{\Xi}_{(k)},
\end{equation}
where
\begin{equation}
\widetilde{\Xi}_{(k)}=\sum_{i=1}^n\widetilde{\xi}_{(k)i}(\mathbf{x},\mathbf{u}_{(0)},\ldots,\mathbf{u}_{(k)})
\frac{\partial}{\partial x_i}
+\sum_{\alpha=1}^m\widetilde{\eta}_{(k)\alpha}(\mathbf{x},\mathbf{u}_{(0)},\ldots,\mathbf{u}_{(k)})\frac{\partial}{\partial u_\alpha}.
\end{equation}

Since we have to deal with differential equations, we need to prolong the Lie generator
to account for the transformation of derivatives. This is done as in classical Lie group analysis of differential equations,
\emph{i.e.}, the derivatives are transformed in such a way the contact conditions are preserved. 
Of course, in the expression of prolongations, we need to take into account the expansions of $\xi_i$, $\eta_\alpha$ and $u_\alpha$,  and drop the $O(\varepsilon^{p+1})$ terms.

\begin{example}
Let $p=1$, and consider the approximate Lie generator
\begin{equation}
\begin{aligned}
\Xi &\approx \sum_{i=1}^n\left(\xi_{(0)i}+\varepsilon\left(
 \xi_{(1)i}+\sum_{\beta=1}^m\frac{\partial \xi_{(0)i}}{\partial u_{(0)\beta}}u_{(1)\beta}\right)\right)\frac{\partial}{\partial x_i}\\
&+\sum_{\alpha=1}^m\left(\eta_{(0)\alpha}+\varepsilon\left(
 \eta_{(1)\alpha}+\sum_{\beta=1}^m\frac{\partial \eta_{(0)\alpha}}{\partial u_{(0)\beta}}u_{(1)\beta}\right)\right)\frac{\partial}{\partial u_\alpha},
 \end{aligned}
\end{equation}
where $\xi_{(0)i}$, $\xi_{(1)i}$, $\eta_{(0)\alpha}$ and $\eta_{(1)\alpha}$ depend on $(\mathbf{x},\mathbf{u}_{(0)})$.
The first order prolongation is
\begin{equation}
\Xi^{(1)}\approx\Xi + \sum_{\alpha=1}^m\sum_{i=1}^n \eta_{\alpha,i}\frac{\partial}{\partial \frac{\partial u_\alpha}{\partial x_i}},
\end{equation}
where
\begin{equation}
\begin{aligned}
\eta_{\alpha,i} &= \frac{D}{D x_i}\left(\eta_{(0)\alpha}+\varepsilon\left(
 \eta_{(1)\alpha}+\sum_{\beta=1}^m\frac{\partial \eta_{(0)\alpha}}{\partial u_{(0)\beta}}u_{(1)\beta}\right)\right)\\
 &-\sum_{j=1}^n \frac{D}{D x_i}\left(\xi_{(0)j}+\varepsilon\left(
 \xi_{(1)j}+\sum_{\beta=1}^m\frac{\partial \xi_{(0)j}}{\partial u_{(0)\beta}}u_{(1)\beta}\right)\right)
 \left(\frac{\partial u_{(0)\alpha}}{\partial x_j}+\varepsilon \frac{\partial u_{(1)\alpha}}{\partial x_j}\right),
\end{aligned}
\end{equation}
along with the Lie derivative
\begin{equation}
\frac{D}{Dx_i}=\frac{\partial}{\partial x_i}+\sum_{\alpha=1}^m \frac{\partial u_{(0)\alpha}}{\partial x_i}
\frac{\partial}{\partial u_{(0)\alpha}}.
\end{equation}
Similar reasonings lead to higher order prolongations.
\end{example}

The approximate (at the order $p$) invariance condition of a differential equation reads:
\begin{equation}
\left.\Xi^{(r)}(\Delta)\right|_{\Delta=O(\varepsilon^{p+1})}= O(\varepsilon^{p+1}).
\end{equation}
In the resulting condition we have to insert the expansion of $\mathbf{u}$ in order to obtain the determining 
equations at the various orders in $\varepsilon$. The integration of the determining equations provides the infinitesimal generators of the admitted approximate symmetries, and approximate invariant solutions may be determined
by solving the invariant surface conditions
\begin{equation}
\mathbf{Q}\equiv \sum_{k=0}^p\varepsilon^k\left(\sum_{i=1}^n\widetilde{\xi}_{(k)i}(\mathbf{x},\mathbf{u}_{(0)},\ldots,\mathbf{u}_{(k)})
\frac{\partial \mathbf{u}}{\partial x_i}-
\widetilde{\boldsymbol \eta}_{(k)}(\mathbf{x},\mathbf{u}_{(0)},\ldots,\mathbf{u}_{(k)})\right)=O(\varepsilon^{p+1}),
\end{equation}
where $\mathbf{u}$ is expanded as in \eqref{expansion_u}.

The Lie generator $\widetilde{\Xi}_{(0)}$ is always a symmetry of the unperturbed equations ($\varepsilon=0$); the  \emph{correction}
terms $\displaystyle\sum_{k=1}^p\varepsilon^k\widetilde{\Xi}_{(k)}$ give
the deformation of the symmetry due to the terms involving $\varepsilon$. 
Not all symmetries of the unperturbed equations are admitted as the zeroth terms of approximate 
symmetries; the symmetries
of the unperturbed equations that are the zeroth terms of approximate symmetries are called \emph{stable 
symmetries} \cite{BGI-1989}.  

\section{Conditional symmetries and approximate conditional symmetries}
\label{sec3}
It is known that many equations of interest in concrete applications possess poor Lie symmetries. More rich reductions leading to wide classes of exact solutions are possible by using the conditional symmetries, which are obtained by appending to the equations at hand some differential constraints. An important class is represented by Q--conditional symmetries, where the constraints to be added to the differential equation \eqref{de} are the invariant surface conditions \eqref{Qconstraint} and their differential consequences. In such a case, Q--conditional symmetries are expressed by generators $\Xi$ such that
\begin{equation}
\left.\Xi^{(r)}(\Delta)\right|_{\mathcal{M}}=0,
\end{equation}
where $\mathcal{M}$ is the manifold of the jet space defined by the system of equations
\begin{equation}
\Delta=0, \qquad \mathbf{Q}=0, \qquad \frac{\partial^{|s|} \mathbf{Q}}{\partial x_1^{s_1}\ldots\partial x_n^{s_n}}=0,
\end{equation}
where $1\le |s|=s_1+\ldots +s_n \le r-1$. 

It is easily proved that a (classical) Lie symmetry is a Q--conditional symmetry. However, differently from Lie symmetries, all possible conditional symmetries of a differential equation form a set which is not a Lie algebra in the general case.
Moreover, if the infinitesimals of a Q--conditional symmetry are multiplied by an arbitrary smooth function $f(\mathbf{x},\mathbf{u})$ we have still a Q--conditional symmetry. This means that we can look for Q--conditional symmetries in $n$ different situations, where $n$ is the number of independent variables, along with the following constraints:
\begin{itemize}
\item $\xi_1=1$; 
\item $\xi_i=1$ and $\xi_{j}=0$ for $1\le j <i \le n$.
\end{itemize}

When $m>1$ (more than one dependent variable), one can look 
\cite{Cherniha-JPA2010,ChernihaDavydovych-MCM2011,ChernihaDavydovych-CNSNS2012}
for Q--conditional symmetries of $q$--th type $(1\le q \le m)$, by requiring the condition
\begin{equation}
\left.\Xi^{(r)}(\Delta)\right|_{\mathcal{M}_q}=0,
\end{equation}
where $\mathcal{M}_q$ is the manifold of the jet space defined by the system of equations
\begin{equation}
\begin{aligned}
&\Delta=0, \qquad Q_{i_1}=\ldots = Q_{i_q}=0, \\ 
&\frac{\partial^{|s|} {Q_{i_1}}}{\partial x_1^{s_1}\ldots\partial x_n^{s_n}}=\ldots = \frac{\partial^{|s|} {Q_{i_q}}}{\partial x_1^{s_1}\ldots\partial x_n^{s_n}}= 0,
\end{aligned}
\end{equation}
where $1\le |s|=s_1+\ldots +s_n \le r-1$, $1\le i_1<\cdots<i_q \le m$.
It is obvious that we may have $\binom{m}{q}$ different Q--conditional symmetries of $q$--th type, whereas for $q=0$ we simply have classical Lie symmetries.

Remarkably, Q--conditional symmetries allow for symmetry reductions of differential equations and provide explicit solutions not always obtainable with classical symmetries.

Now we have all the elements to define approximate Q--conditional symmetries. 

\begin{definition}[Approximate Q--conditional symmetries]
Given the differential equation
\begin{equation}
\Delta\approx \sum_{k=0}^p\varepsilon^k\widetilde{\Delta}_{(k)}\left(\mathbf{x},\mathbf{u}_{(0)},\mathbf{u}^{(r)}_{(0)},
\ldots,\mathbf{u}_{(k)},\mathbf{u}^{(r)}_{(k)}\right)=0.
\end{equation}
and the approximate invariant surface condition
\begin{equation}
\mathbf{Q}\equiv \sum_{k=0}^p\varepsilon^k\left(\sum_{i=1}^n\widetilde{\xi}_{(k)i}(\mathbf{x},\mathbf{u}_{(0)},\ldots,\mathbf{u}_{(k)})
\frac{\partial \mathbf{u}}{\partial x_i}-
\widetilde{\boldsymbol \eta}_{(k)}(\mathbf{x},\mathbf{u}_{(0)},\ldots,\mathbf{u}_{(k)})\right)=O(\varepsilon^{p+1}),
\end{equation}
the approximate Q--conditional symmetries of order $p$ are found by requiring
\begin{equation}
\left.\Xi^{(r)}(\Delta)\right|_{\mathcal{M}}=O(\varepsilon^{p+1}),
\end{equation}
where $\mathcal{M}$ is the manifold of the jet space defined by the system of equations
\begin{equation}
\Delta=O(\varepsilon^{p+1}), \qquad \mathbf{Q}=O(\varepsilon^{p+1}), \qquad \frac{\partial^{|s|} \mathbf{Q}}{\partial x_1^{s_1}\ldots\partial x_n^{s_n}}=O(\varepsilon^{p+1}),
\end{equation}
where $1\le |s|=s_1+\ldots +s_n \le r-1$.
\end{definition}

When $m>1$, \emph{i.e.}, differential equations involving more than one dependent variable, approximate Q--conditional symmetries of $q$--th type may be defined.

\begin{definition}[Approximate Q--conditional symmetries of $q$--th type]
Given the differential equation
\begin{equation}
\Delta\approx \sum_{k=0}^p\varepsilon^k\widetilde{\Delta}_{(k)}\left(\mathbf{x},\mathbf{u}_{(0)},\mathbf{u}^{(r)}_{(0)},
\ldots,\mathbf{u}_{(k)},\mathbf{u}^{(r)}_{(k)}\right)=0.
\end{equation}
and the approximate invariant surface condition
\begin{equation}
\mathbf{Q}\equiv \sum_{k=0}^p\varepsilon^k\left(\sum_{i=1}^n\widetilde{\xi}_{(k)i}(\mathbf{x},\mathbf{u}_{(0)},\ldots,\mathbf{u}_{(k)})
\frac{\partial \mathbf{u}}{\partial x_i}-
\widetilde{\boldsymbol \eta}_{(k)}(\mathbf{x},\mathbf{u}_{(0)},\ldots,\mathbf{u}_{(k)})\right)=O(\varepsilon^{p+1}),
\end{equation}
the approximate Q--conditional symmetries of $q$--th type of order $p$ are found by requiring
\begin{equation}
\left.\Xi^{(r)}(\Delta)\right|_{\mathcal{M}_q}=O(\varepsilon^{p+1}),
\end{equation}
where $\mathcal{M}_q$ is the manifold of the jet space defined by the system of equations
\begin{equation}
\begin{aligned}
&\Delta=O(\varepsilon^{p+1}), \qquad Q_{i_1}=\ldots = Q_{i_q}=O(\varepsilon^{p+1}), \\ 
&\frac{\partial^{|s|} {Q_{i_1}}}{\partial x_1^{s_1}\ldots\partial x_n^{s_n}}=\ldots = \frac{\partial^{|s|} {Q_{i_q}}}{\partial x_1^{s_1}\ldots\partial x_n^{s_n}}= O(\varepsilon^{p+1}),
\end{aligned}
\end{equation}
where $1\le |s|=s_1+\ldots +s_n \le r-1$, $1\le i_1<\cdots<i_q \le m$.
\end{definition}

For simplicity, hereafter we will consider  the case $p=1$; for higher values of $p$ things are  similar but at the cost of an increased amount of computations.

Similarly to the case of exact Q--conditional symmetries, we can look for approximate  Q--conditional symmetries in $n$ different situations, where $n$ is the number of independent variables, along with the constraints
\begin{itemize}
\item $\xi_{(0)1}=1$, $\xi_{(k)1}=0$ for $k=1,\ldots,p$; 
\item $\xi_{(0)i}=1$,  $\xi_{(k)i}=0$, $\xi_{(0)j}=\xi_{(k)j}=0$ for $1\le j <i \le n$ and $k=1,\ldots,p$.
\end{itemize}

\section{Application}
\label{sec4}
In this Section, we consider the equation
\begin{equation}
\label{RDChyp}
\varepsilon u_{tt}+u_{t}-(uu_x)_x-\alpha u u_x+\beta u(1-\gamma u)=0,
\end{equation}
where $\varepsilon$ is a small constant parameter,
that represents a hyperbolic version of the reaction--diffusion--convection equation \cite{Murray}
\begin{equation}
\label{RDC}
u_{t}-(uu_x)_x-\alpha u u_x+\beta u(1-\gamma u)=0,
\end{equation}
$\alpha$, $\beta$ and $\gamma$ being real positive constants.

In \cite{Plyukhin}, the Q--conditional symmetries of a family of partial differential equations containing \eqref{RDC} and some exact solutions have been explicitly determined. Equation \eqref{RDC}, supplemented by the invariant surface condition (and its first order differential consequences)
\begin{equation}
\label{Qcond}
u_t+\xi(t,x,u)u_x-\eta(t,x,u)=0,
\end{equation}
admits the following generators of Q--conditional symmetries:
\begin{equation}
\label{Qoper}
\begin{aligned}
&\Xi_1=\frac{\partial}{\partial t}+f(t)\frac{\partial}{\partial x}+\frac{f^\prime(t)}{f(t)}u\frac{\partial}{\partial u},\\
&\Xi_2=\frac{\partial}{\partial t}+g(t)u\frac{\partial}{\partial u},
\end{aligned}
\end{equation}
where $f(t)$ and $g(t)$ are functions of time satisfying the following differential equations
\begin{equation}
\label{constr_fg}
\begin{aligned}
&f f^{\prime\prime}-2f^{\prime 2}-\beta f f^\prime=0,\\
&g^\prime -\beta g-g^2 =0,
\end{aligned}
\end{equation}
whose integration provides
\begin{equation}
\label{fg}
\begin{aligned}
&f=\frac{k_1}{1-k_2\exp(\beta t)},\\
&g=\beta\frac{k_1\exp(\beta t)}{1-k_1\exp(\beta t)}, \quad\hbox{or}\quad g=-\beta,
\end{aligned}
\end{equation}
with $k_1$ and $k_2$ arbitrary constants.

Here, we limit ourselves to report the Q--conditional invariant solutions of \eqref{RDC} corresponding to operator $\Xi_2$ in \eqref{Qoper} along 
with the assumption $g=-\beta$. By solving together the invariant surface condition and \eqref{RDC}, according to the values of the parameters $\alpha$, $\beta$ and $\gamma$, after some straightforward algebra, 
three exact solutions are determined \cite{Plyukhin}:
\begin{itemize}
\item $8\beta\gamma-\alpha^2=\delta^2$:   
\begin{equation}
\label{sol1RDC}
u(t,x)=c_1\exp\left(-\beta t-\frac{\alpha}{4} x\right)\sqrt{\cos\left(\frac{\delta}{2}x+c_2\right)};
\end{equation}
\item $8\beta\gamma-\alpha^2=0$:
\begin{equation}
\label{sol2RDC}
u(t,x)=c_1\exp\left(-\beta t-\frac{\alpha}{4}x\right)\sqrt{x+c_2};
\end{equation}
\item $8\beta\gamma-\alpha^2=-\delta^2$:
\begin{equation}
\label{sol3RDC}
u(t,x)=c_1\exp\left(-\beta t-\frac{\alpha+\delta}{4}x\right)\sqrt{\exp\left(\delta x\right)+c_2},
\end{equation}
\end{itemize}
$c_1$ and $c_2$ being arbitrary constants.

Solutions \eqref{sol1RDC}, \eqref{sol2RDC} and \eqref{sol3RDC} have been already determined in \cite{Plyukhin}, and are reported here
just to compare them with the approximate Q--conditional invariant solutions of equation \eqref{RDChyp}.  

Now, let us consider equation \eqref{RDChyp} and look for the admitted first order approximate Q--conditional symmetries generated by the infinitesimal operator
\begin{equation}
\begin{aligned}
\Xi&=\frac{\partial}{\partial t}+\xi(t,x,u;\varepsilon)\frac{\partial}{\partial x}+\eta(t,x,u;\varepsilon)\frac{\partial}{\partial u}\approx\\ 
&\approx\frac{\partial}{\partial t}+\left(\xi_{(0)}+\varepsilon\left(\xi_{(1)}+\frac{\partial \xi_{(0)}}{\partial u_{(0)}}u_{(1)}\right)\right)\frac{\partial}{\partial x}\\&+\left(\eta_{(0)}+\varepsilon\left(
\eta_{(1)}+\frac{\partial \eta_{(0)}}{\partial u_{(0)}}u_{(1)}\right)\right)\frac{\partial}{\partial u}.
\end{aligned}
\end{equation}

The required computations for determining the infinitesimals of the approximate Q--conditional symmetries are done immediately by using the \textsc{Reduce} \cite{Reduce} package \textsc{ReLie} \cite{Oliveri:relie}. 

Here, we restrict to consider a subclass of first order approximate Q--conditional symmetries;
the complete classification of the admitted approximate Q--conditional symmetries and the corresponding approximate solutions to equation \eqref{RDChyp} will be the object of  a paper under preparation. 

Expanding $u(t,x;\varepsilon)$ at first order in $\varepsilon$, 
\begin{equation}
u(t,x;\varepsilon)=u_{(0)}(t,x)+\varepsilon u_{(1)}(t,x)+O(\varepsilon^2),
\end{equation}
let us consider the approximate Q--conditional symmetries that are first order perturbations of the Q--conditional symmetry generated by
\begin{equation}
\Xi= \frac{\partial}{\partial t}-\beta u_{(0)}\frac{\partial}{\partial u_{(0)}}.
\end{equation}

By solving the determining equations, we have the approximate generator
\begin{equation}
\label{ApproxOper}
\begin{aligned}
\Xi&\approx  \frac{\partial}{\partial t}+\varepsilon\left(\frac{1}{2}\exp(\beta t)f_2(x)u_{(0)}^2-\exp(-\beta t)f_3(x)\right)\frac{\partial}{\partial x}\\
&-\left(\beta u_{(0)}+\varepsilon
\left(\beta u_{(1)} -\exp(-3\beta t)f_1(x)-\frac{\exp(\beta t)}{4}\left(f_2^\prime(x) -\alpha f_2(x)\right)u_{(0)}^3\right.\right.\\
&\left.\left.+\frac{\exp(-\beta t)}{4}\left(f_3^\prime(x)-\alpha f_3(x)\right) u_{(0)}-\left(\frac{k_1\exp(-\beta t)}{4}-\beta^2\right)u_{(0)}\right)\right)
\frac{\partial}{\partial u},
\end{aligned}
\end{equation}
with $f_1$, $f_2$ and $f_3$ functions of $x$ subjected to the constraints:
\begin{equation}
\begin{aligned}
&f_1^{\prime\prime}+\alpha f_1^\prime+2\beta\gamma f_1=0,\\
&f_2^{\prime\prime}-\alpha f_1^\prime+2\beta\gamma f_2=0,\\
&f_3^{\prime\prime}-(\alpha^2 -8\beta\gamma) f_3+k_2=0,
\end{aligned}
\end{equation}
and $k_1$, $k_2$ arbitrary constants.

Three cases have to be distinguished depending on the values of the coefficients $\alpha$, $\beta$ and $\gamma$:
\begin{itemize}
\item $\alpha^2-8\beta\gamma=\delta^2$:   
\begin{equation}
\begin{aligned}
&f_1=k_3 \exp\left(-\frac{\alpha+\delta}{2}x\right)+k_4\exp\left(-\frac{\alpha-\delta}{2}x\right),\\
&f_2=k_5 \exp\left(\frac{\alpha+\delta}{2}x\right)+k_6\exp\left(\frac{\alpha-\delta}{2}x\right),\\
&f_3=k_7\exp(\delta x)+k_8\exp(-\delta x)+\frac{k_2}{\delta^2};
\end{aligned}
\end{equation}

\item $\alpha^2-8\beta\gamma=0$:   
\begin{equation}
\begin{aligned}
&f_1=\exp\left(-\frac{\alpha}{2}x\right)(k_3 +k_4x),\\
&f_2=\exp\left(\frac{\alpha}{2}x\right)(k_5+k_6x),\\
&f_3=-\frac{k_2}{2}x^2+k_7x+k_8;
\end{aligned}
\end{equation}
\item $\alpha^2-8\beta\gamma=-\delta^2$:   
\begin{equation}
\begin{aligned}
&f_1=\exp\left(-\frac{\alpha}{2}x\right)\left(k_3 \cos\left(\frac{\delta}{2}x\right)+k_4\sin\left(\frac{\delta}{2}x\right)\right),\\
&f_2=\exp\left(\frac{\alpha}{2}x\right)\left(k_3 \cos\left(\frac{\delta}{2}x\right)+k_4\sin\left(\frac{\delta}{2}x\right)\right),\\
&f_3=k_7\cos(\delta x)+k_8\sin(\delta x)-\frac{k_2}{\delta^2},
\end{aligned}
\end{equation}
\end{itemize}
where $k_3,\ldots,k_8$ are arbitrary constants.

\begin{figure}
\centering
\includegraphics[width=0.8\textwidth]{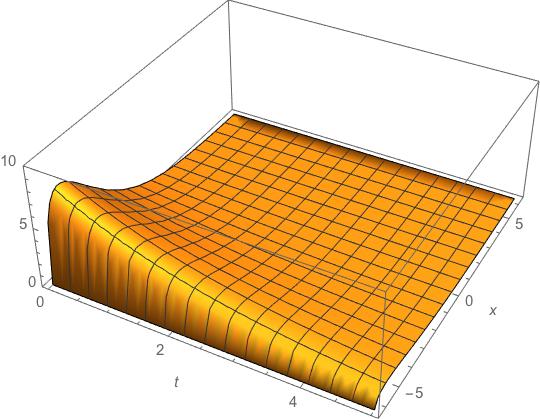}
\caption{\label{fig1}Plot of solution \eqref{sol1} with the parameters $\varepsilon= 0.03$, $\alpha=2$, 
$\beta=0.4$, $\gamma=1.33$, $c_1= 1$, $c_2= 0$, $c_3=1$, $c_4=0$.}
\end{figure}
\begin{figure}
\centering
\includegraphics[width=0.8\textwidth]{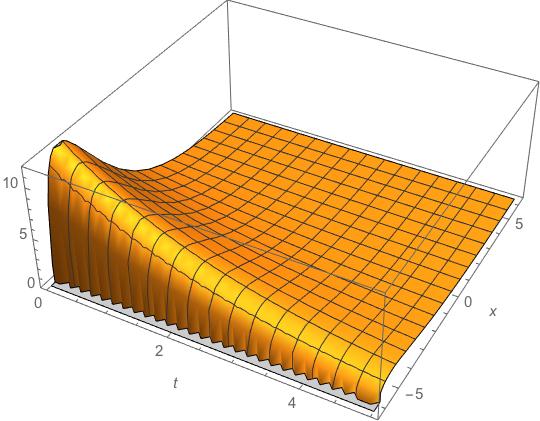}
\caption{\label{fig2}Plot of solution \eqref{sol2} with the parameters $\varepsilon= 0.03$, $\alpha=2$, 
$\beta=0.4$, $\gamma=1.25$, $c_1= 1$, $c_2= 2\pi$, $c_3=1$, $c_4=0$.}
\end{figure}
\begin{figure}
\centering
\includegraphics[width=0.8\textwidth]{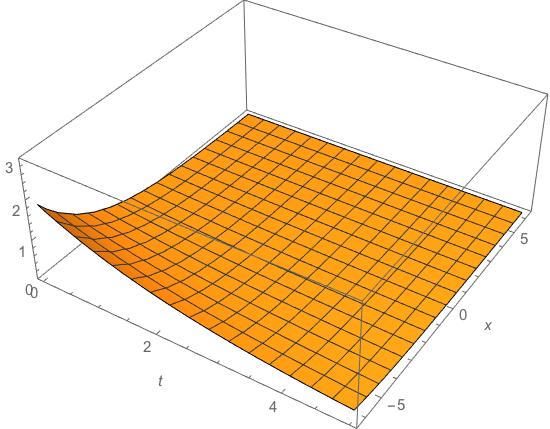}
\caption{\label{fig3}Plot of solution \eqref{sol3} with the parameters $\varepsilon= 0.03$, $\alpha=2$, 
$\beta=0.4$, $\gamma=1.17$, $c_1= 1$, $c_2= 0$, $c_3=1$, $c_4=0$.}
\end{figure}

By considering the approximate generator 
\begin{equation}
\Xi\approx \frac{\partial}{\partial t}-\left(\beta u_{(0)}+\varepsilon (\beta^2 u_{(0)}+\beta u_{(1)})\right)\frac{\partial}{\partial u},
\end{equation}
(obtained by choosing $f_1=f_2=f_3=k_1=k_2=0$ in \eqref{ApproxOper}), after some algebra, we are able to find the following approximate solutions of equation \eqref{RDChyp} that, for $\varepsilon=0$ are those reported above for the reaction--diffusion--convection equation \eqref{RDC}:
\begin{itemize}
\item $8\beta\gamma-\alpha^2=\delta^2$:   
\begin{equation}
\label{sol1}
\begin{aligned}
u(t,x;\varepsilon)&=\exp\left(-\beta t-\frac{\alpha}{4} x\right)\sqrt{\cos\left(\frac{\delta}{2}x+c_2\right)}\times\\
&\times \left(c_1+\varepsilon \left(c_3\frac{\cos\left(\frac{\delta}{2} x+c_4\right)}{\cos\left(\frac{\delta}{2} x+c_2\right)}
-c_1\beta^2 t \right)\right);
\end{aligned}
\end{equation}
\item $8\beta\gamma-\alpha^2=0$:
\begin{equation}
\label{sol2}
\begin{aligned}
u(t,x;\varepsilon)&=\exp\left(-\beta t-\frac{\alpha}{4}x\right)\sqrt{x+c_2}\times\\
&\times\left(c_1+\varepsilon\left(\frac{c_3 x+c_4}{x+c_2}-c_1\beta^2 t\right)\right);
\end{aligned}
\end{equation}
\item $8\beta\gamma-\alpha^2=-\delta^2$:
\begin{equation}
\label{sol3}
\begin{aligned}
u(t,x;\varepsilon)&=\exp\left(-\beta t-\frac{\alpha+\delta}{4}x\right)\sqrt{\exp\left(\delta x\right)+c_2}\times\\
&\times\left(c_1+\varepsilon\left(\frac{c_3\exp\left(\delta x\right)+c_4\delta}{\delta\exp\left(\delta x\right)+c_2}
-c_1\beta^2t\right)\right),
\end{aligned}
\end{equation}
with $c_i$ ($i=1,\ldots,4$) arbitrary constants.
\end{itemize}

Plots of these solutions are displayed in Figures~\ref{fig1}--\ref{fig3}.

By using different Q--conditional symmetries, other classes of approximate solutions can be explicitly obtained. Here, we report only a few; a more complete list will be contained in a forthcoming paper. 

For instance, by using the approximate Q--conditional symmetry  generated by
\begin{equation}
\begin{aligned}
\Xi&\approx \frac{\partial}{\partial t}-\varepsilon \exp(-\beta t-\delta x)\frac{\partial}{\partial x}\\
&+\left(-\beta u_{(0)}+
\varepsilon\left(\left(\frac{\alpha+\delta}{4}\exp(-\beta t-\delta x)-\beta^2\right)u_{(0)}-\beta u_{(1)}\right)\right)\frac{\partial}{\partial u},
\end{aligned}
\end{equation}
when $\alpha^2-8\beta\gamma=\delta^2$,  we get the approximate solution
\begin{equation*}
\begin{aligned}
&u(t,x;\varepsilon)\approx \exp(-\beta t)\left(c_1\exp\left(-\frac{\alpha-\delta}{4}x\right)+\varepsilon\left(\frac{16}{c_1^2(\alpha-\delta)(\alpha+3\delta)}\right.\right.\\
&\left.\left.\qquad+\left(c_2-c_1\beta^ 2 t\right)\exp\left(-\frac{\alpha+\delta}{4}x\right)+
\left(c_3-\frac{c_1\delta}{2\beta}\exp(-\beta t)\right)\exp\left(-\frac{\alpha+3\delta}{4}x\right)\right)\right).
\end{aligned}
\end{equation*}
When $\alpha^2-8\beta\gamma=0$, by using the approximate Q--conditional symmetry generated by
\begin{equation}
\begin{aligned}
\Xi&\approx \frac{\partial}{\partial t}-\varepsilon \exp(-\beta t)\frac{\partial}{\partial x}\\
&+\left(-\beta u_{(0)}+
\varepsilon\left(\left(\frac{\alpha}{4}\exp(-\beta t)-\beta^2\right)u_{(0)}-\beta u_{(1)}\right)\right)\frac{\partial}{\partial u},
\end{aligned}
\end{equation}
we obtain the approximate solution
\begin{equation*}
\begin{aligned}
&u(t,x;\varepsilon)\approx \exp(-\beta t)\left(c_1\exp\left(-\frac{\alpha}{4}x\right)\sqrt{x}+
\varepsilon\left(-\frac{2}{\alpha} +\frac{2}{\sqrt{\alpha x}}\left(\frac{2}{\alpha}+x\right)\mathcal{D}\left(\frac{\sqrt{\alpha x}}{2}\right)\right.\right.\\
&\left.\left.\qquad+\left(\frac{1}{\sqrt{x}}\left(c_2-\frac{c_1\exp(-\beta t)}{2\beta}\right)+\sqrt{x}(c_3-c_1\beta t^2)\exp\left(-\frac{\alpha x}{4}\right)\right)\right)\right),
\end{aligned}
\end{equation*}
where
\begin{equation}
\mathcal{D}\left(\frac{\sqrt{\alpha x}}{2}\right)=\exp\left(-\frac{\alpha x}{4}\right)\int_{0}^{\sqrt{\alpha x}/2}\exp(-z^2)\, dz
\end{equation}
is the Dawson function.
\subsection*{Concluding remarks}
\label{sec5}
In this paper, by adopting a recently proposed approach to approximate Lie symmetries of differential equations involving \emph{small terms}, 
we have given a consistent definition of approximate Q--conditional symmetries. The theoretical framework has been applied to a partial differential equation that represents a hyperbolic version of a reaction--diffusion--convection equation \cite{Murray}.
We obtained some classes of approximate Q--conditional invariant solutions. The complete classification of approximate Q--conditional symmetries, as well as the derivation of the corresponding approximate solutions, will be contained in a forthcoming paper.

\subsection*{Acknowledgments}
Work supported by GNFM of ``Istituto Nazionale di Alta Matematica''.

\end{document}